\RequirePackage{fix-cm}
\documentclass[twocolumn]{svjour3}          
\smartqed  
\usepackage{graphicx}
\usepackage{natbib}  
\begin{document}
\title{The potential for complex computational models of aging \thanks{$^*$ Corresponding author} }
\author{Spencer Farrell$^a$ \and Garrett Stubbings$^a$ \and Kenneth Rockwood$^b$ \and Arnold Mitnitski$^b$ \and Andrew Rutenberg$^{a*}$}

\authorrunning{Farrell et al} 

\institute{$^a$               Department of Physics and Atmospheric Science,
              Dalhousie University, Halifax,
              Nova Scotia, Canada B3H 4R2
           \and    \\       
           $^b$ Division of Geriatric Medicine, Dalhousie University, Halifax, Nova Scotia, Canada B3H 2E1 \\
           Email addresses: spencer.farrell@dal.ca (S. Farrell), \\
                adr@dal.ca (A. Rutenberg).
}
\date{Received: date / Accepted: date} 
\maketitle

\begin{abstract}
The gradual accumulation of damage and dysregulation during the aging of living organisms can be quantified. Even so, the aging process is complex and has multiple interacting physiological scales -- from the molecular to cellular to whole tissues. In the face of this complexity, we can significantly advance our understanding of aging with the use of computational models that simulate realistic individual trajectories of health as well as mortality. To do so, they must be systems-level models that incorporate interactions between measurable aspects of age-associated changes. To incorporate individual variability in the aging process, models must be stochastic. To be useful they should also be predictive, and so must be fit or parameterized by data from large populations of aging individuals. In this perspective, we outline where we have been, where we are, and where we hope to go with such computational models of aging. Our focus is on data-driven systems-level models, and on their great potential in aging research.
\end{abstract}

\section{Introduction: Challenges of Studying Aging}
Computational models are essential to make state-of-the-art predictions or to understand mechanisms within complex non-linear, stochastic, and interconnected systems such as the economy, the weather, or the climate. In this section we outline how aging organisms also represent complex, interconnected dynamical systems that are challenging to study. 

Aging populations exhibit increasing mortality rates. For humans, the risk of dying increases approximately exponentially for older ages -- the famous Gompertz law of mortality \citep{Kirkwood2015}. Before death, individual health can be assessed and summarized in many ways. One such measure is provided by the Frailty Index (FI) which is the proportion of ``things wrong'' from a large selection of possible age-related deficits of health and function \citep{Mitnitski2001}. The FI is robust, flexible, and is strongly correlated with various outcome measures including mortality \citep{Rockwood2005, Evans2014}. Alternatively, Biological Age (BA) is an ``effective age'' defined in terms of an individual's health, often using molecular aspects of health such as epigenetic methylation \citep{Hannum2013, Horvath2013, Levine2020}. Other summary measures of health have been developed, including allostatic load \citep{McEwen1993}, and physiological disregulation \citep{Milot2014}. Different summary measures of health are not necessarily strongly correlated with each other at the individual level \citep{Li2020}, indicating that aging is a multi-dimensional process.

As assessed by the FI, the distribution of health measures broadens with age, corresponding to distinctive individual trajectories of health \citep{Rockwood2004}. Worsening health over an individual's life is a random process, and is described as a stochastic accumulation of damage. It is thought that this damage underlies the increased mortality with age that is characterized by Gompertz's law \citep{Gavrilov2001}.

Remarkably, even relatively simple empirical observations about aging and mortality are not well understood. Resolving the question of whether Gompertz's law applies for extremely old populations or whether they exhibit a mortality deceleration or plateau remains challenging due to small data-sets  \citep{Gavrilov2019}. The `mortality-morbidity paradox' whereby female populations live longer than male populations, despite male populations apparently having better health, remains largely unexplained \citep{Gordon2018, Gordon2017, Kulminski2008}. The mechanisms behind historically changing health and mortality within national populations \citep{Crimmins2015, Colchero2016}, or behind differences between different socio-economic groups within a population \citep{Andrew2012}, are difficult to assess. 

We see four big questions in aging research: how can we better observe health and mortality across large populations, how can we better understand the mechanisms or causes underlying what we observe, how can we better predict outcomes at an individual or population level, and, finally, how can we better intervene to decrease mortality and to improve health during aging? The challenges implicit in addressing these questions are interconnected: progress in any of these directions will support and direct progress in the others.

Success in aging research crucially depends on the broad availability of high-quality data. National studies, especially those that include longitudinal data on study participants, such as the CSHA \citep{CSHA1994}, CLSA \citep{Raina2009}, NHANES \citep{NHANES2014}, BLSA \citep{Ferrucci2008}, ELSA \citep{Steptoe2014}, and the UK Biobank \citep{Sudlow2015}, are of increasing importance and utility. Emerging sources of ``big''-data include electronic health records (EHR) \citep{Clegg2016}, molecular 'omics, and individual telemetry provided by health monitors or cellphones. How and what can we learn from these new sources of aging data?

There are also many scales of health measures to consider: from molecular and cellular, to tissue, to organismal -- including functional or social aspects of the organism. For example, at the molecular scale methylation clocks have emerged as convenient epigenetic hallmarks of health and aging \citep{Hannum2013, Horvath2013}. Other high-throughput technologies such as genomics, transcriptomics, proteomics, meta\-bol\-omics, and microbiomics provide ways of measuring large amounts of data in aging studies \citep{Livshits2018, Lehallier2019, Ahadi2020}. Conversely, clinically relevant aspects of health such as activities of daily living (ADL) and other measures of functional disability are particularly important to the aging of individual older adults. Such `higher' levels of function are dependent on many aspects of `lower-level' biological and molecular function in a variety of tissues. How can we relate and usefully combine observations at different scales?

Continuing the historical advances in either life expectancy or healthy-aging will be increasingly challenging but, naturally, is of great interest to the geroscience community. Targeted interventions for aging individuals will often be in the context of significant comorbidities or polypharmacy. Systemic treatments such as exercise \citep{Fried2016, Partridge2018}, caloric restriction \citep{Most2017, Mattson2017}, or senolytics \citep{Xu2018}, act at cellular or molecular scales but the desired effects are often at the organismal scale. Understanding how  different scales of organismal function interact with each other as a dynamical system should help us to effectively translate advances from one scale to another, and identify interventions that target the lower-level biological function before it manifests as functional disability \citep{Ferrucci2018}. Optimal interventions will likely depend on the health-state and age of the individual. How can we best provide individualized medicine for aging individuals?

Animal models of aging have instructive similarities and differences with respect to human aging \citep{Cohen2018}. Furthermore, efficient automated image analysis is starting to lead to high-quality longitudinal studies of model organisms (see e.g. worms \citep{Swierczek2011, Zhang2016}, or flies \citep{Seroude2002}). Developing insights from frequently-measured high-dimen\-sion\-al organismal health states within large model populations will require new suites of analysis and modelling tools, but should yield deeper understanding of the aging process. Simple animal models are particularly amen\-able to studying the multi-scale effects of controlled interventions.

\begin{table*}[t!] 
\caption{Some Challenges and Promise of Complex Computational Models of Aging}
\begin{tabular}{|l|l|}
\hline
{\bf Challenges} & {\bf Promise} \\[2pt]
\hline
Using large heterogeneous and sparse longitudinal datasets  &   Better individualized predictions \\
 & \, \, \, (using E-health and self-reported data)\\[5pt] 
Making population context explicit & Better population health \\
                                    & Better synthetic populations \\[5pt]
Including multiple scales   & Better translation of lab-based interventions \\ 
\, \, \, (e.g. cellular to functional) &  \\
Understanding mechanisms  & \\[5pt]
Predicting and understanding effects of     &  Personalized treatment   \\
\, \, \,       health interventions        & Improved treatment of comorbidities \\[2pt]
\hline
\end{tabular}
\label{Challenges}
\end{table*}

Given the complexity of the aging process, how can computational models make use of available and emerging sources of data in order to improve our understanding of aging within and between populations or species, to better predict individual aging outcomes, and to both understand existing interventions and to develop better individualized interventions in the aging process?  Table~\ref{Challenges} summarizes some of the particular challenges facing us in developing computational models of aging organisms, together with the potential benefits of meeting these challenges. We will return to these challenges in more detail at the end of this perspective. 

\section{Theoretical approaches to Aging} 
Conceptual models such as the hallmarks of aging \citep{Lopez-Otin2013}, seven pillars of aging \citep{Kennedy2014}, or damage accumulation \citep{Kirkwood2005} can provide powerful frameworks for discussion or interpretation of quantitative results, but they are not typically quantitative themselves. While they organize how we think about aging, they do not directly help us to quantitatively characterize observed data or to make quantitative predictions. 

The lightest quantitative models are largely descriptive, or phenomenological. An example is Gompertz's exponential increase of mortality with age for older adults \citep{Kirkwood2015}. This is not an exact model over any age range, but it is a useful approximation. Related to that is the Strehler-Mildvan correlation between the amplitude and exponent of the Gompertz law \citep{Tarkhov2017}, or the discussion of maximum human lifespan \citep{Dolgin2018}. Such phenomenological models beg explanation of mechanisms, but typically do not provide or require that explanation themselves. Furthermore, such simple quantitative models are not able to describe multiple organismal scales of aging. For example, they describe mortality but not health. As a result, they are not typically useful for predicting individual health outcomes.

Predictive models of aging make stronger quantitative assumptions. An example is the proportional hazards model \citep{Cox:1972}. A review of mortality models is provided by \citet{Yashin2000}. Yashin and colleagues also developed the quadratic hazards model (or stochastic process model, SPM) of aging \citep{Yashin2012, Arbeev2016}. It assumes that individual deviations from age-dependent norms of physiological measures interact with and exacerbate each other.  The SPM is an example of a dynamical model of aging.

\subsection{Dynamical Models of Aging}
Dynamical models can be used to explicitly simulate individual health trajectories vs age, i.e. longitudinal data. Because they can generate synthetic individual health data, they can serve much the same role as model organisms -- whereby differences with respect to human aging can be significant but are hopefully  also informative. 

Dynamical models can build in explicit interpretable ``mechanisms''. These do not necessarily need to be fundamental biological mechanisms, but can be ``effective'' mechanisms relating  variables that interact indirectly. Any  success of a dynamical model in reproducing real-world phenomena then suggests the viability of these underlying mechanisms, and allows the modeller to explore other phenomena that arise from the same mechanisms. While such mechanisms cannot be immediately taken as real, the model can be used to identify further experiments and data to test  predictions arising from the mechanisms.

The concept of interactions between health measures can be embodied in a network of pair-wise interactions, where nodes (or ``vertices'') are the health measures while the interactions are connections between nodes (links, or ``edges''). These interactions can be general, and not just the multiplicative interactions usually considered in regression models. Such explicit networks were already used to model mortality \citep{Gavrilov2001, Vural2014}. We developed what we now call a generic network model (GNM) to also model health measures such as the FI, and found that we could describe both population-level aging and mortality with a simple network of interactions that could be implemented on a computer to generate large synthetic populations \citep{Taneja2016, Farrell2016, Farrell2018}. 

\begin{figure}[t!]  
  \center{\includegraphics[width=0.45 \textwidth,trim=5mm 0mm 5mm 0mm, clip]{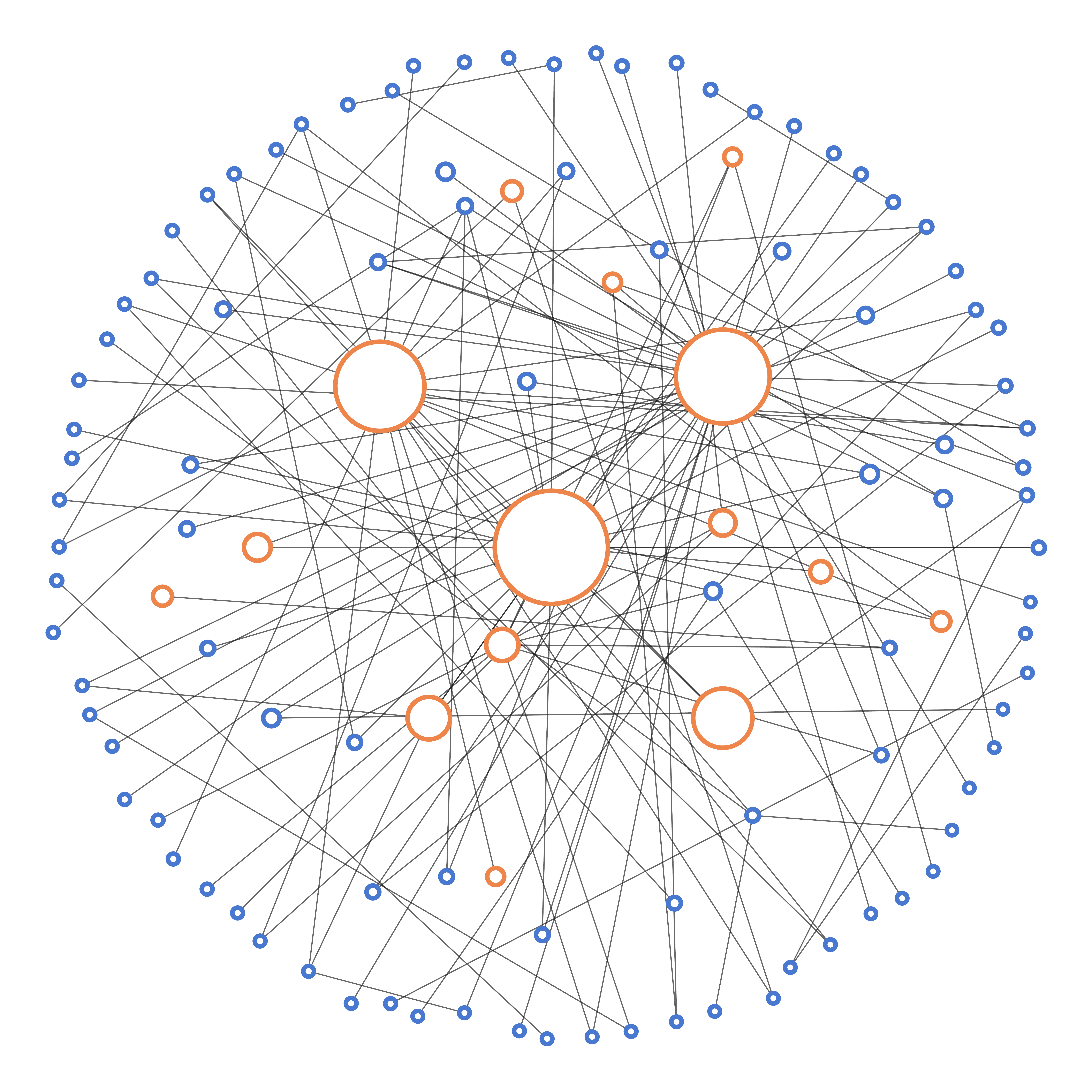}}
  \caption{Links (grey lines) between nodes (circles) in a generic network model (GNM) of aging. Shown are a selection of $112$ nodes, out of $10^4$ used in the model. Nodes do not represent particular observed health variables, but larger circles indicate nodes with more links. The most connected nodes are coloured orange, and are used in the aggregate health measure (frailty index, or FI) of the GNM. Note that the links do not change with age, while the binary health-states of the nodes do change with age between healthy and unhealthy. This model, with stochastic damage rates that can be implemented computationally \citep{Taneja2016, Farrell2016, Farrell2018}, can generate large synthetic populations that reflect observed population-average age-dependent health and mortality rates.}
  \label{fig:undirected}
\end{figure}

The health attributes (nodes) of our GNM did not directly correspond to specific observed health attributes. The reason for that approach was simplicity: all connected nodes  have similar interactions, with simple undirected connections of equal weight, as illustrated in Figure~1. This enabled us to capture population-level health and mortality with only a few parameters. Such generic networked models are useful for conceptual explorations of aging, and have been used to understand mortality curves in different species \citep{Vural2014, Stroustrup2016}, to explore the mutual information between health states and mortality \citep{Farrell2018}, and to explore how to increase longevity through optimized maintenance \citep{Sun2020}. Nevertheless, they are not able to capture realistic individual health states or model their detailed trajectories with age.

To be able to predict detailed individual health states, we would need to empirically capture the many distinct interactions between observed individual health attributes within a population. Reconstructing interactions from observed data is a daunting prospect. If hundreds of individual health attributes are measured, then there are tens of thousands of interactions to determine between all possible pairs of attributes. Initial progress is already being made on smaller-sized problems. Using only cross-sectional binarized health data, we have developed a network model that includes specific observed health attributes \citep{Farrell2020}. To accomplish this, the model parameters are distinct for each node and the network connections between nodes are distinctly weighted. An illustration of such a ``weighted'' network model (WNM) is shown in Figure~2 for 10 nodes.  

\begin{figure}[t!]  
  \center{\includegraphics[width=0.3 \textwidth,trim=5mm 0mm 5mm 0mm, clip]{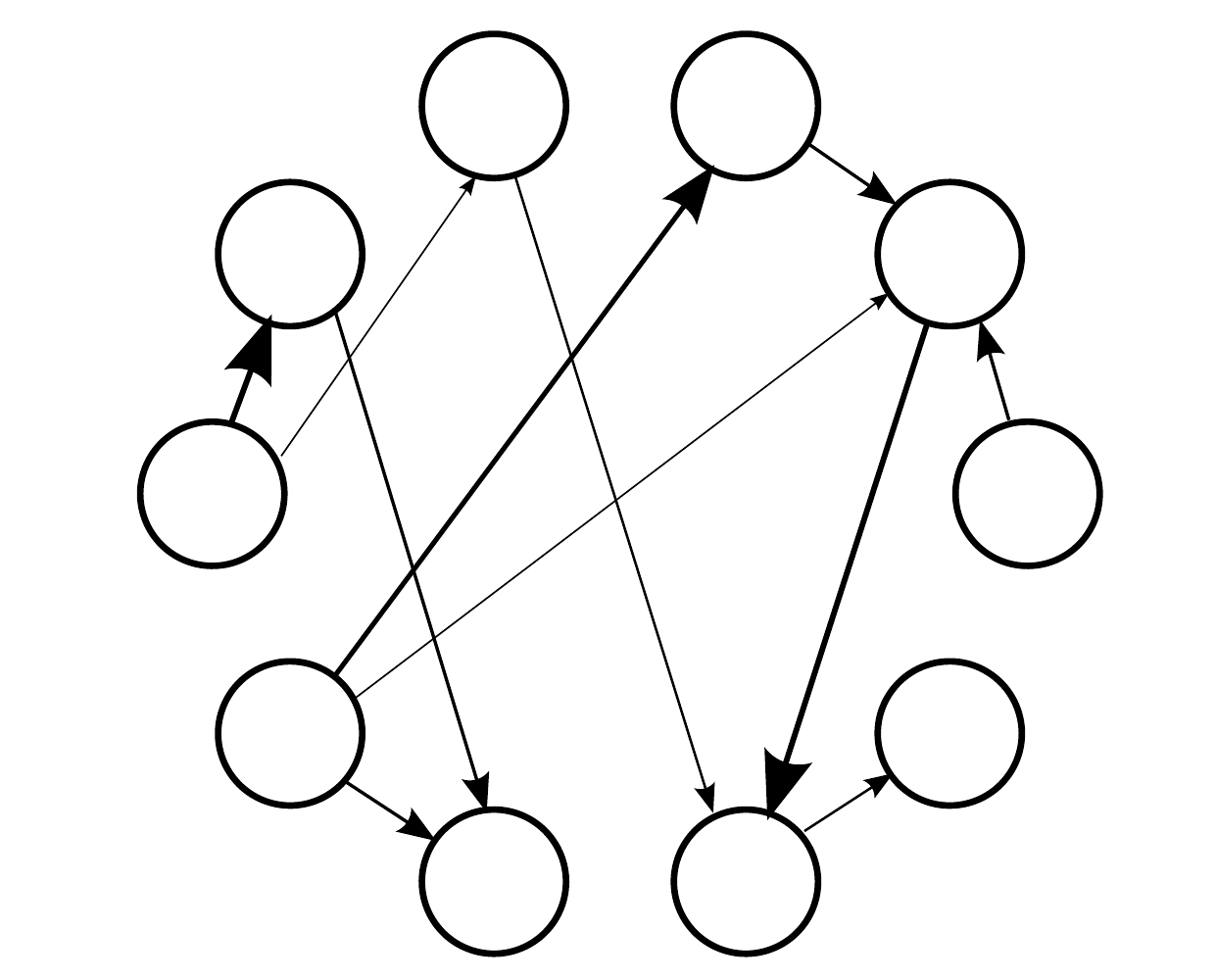}}
  \caption{Weighted and directed links (arrows) between nodes (circles) in a weighted network model (WNM) of aging. Shown are 10 nodes, each of which represents specific observed binary health variables from cross-sectional studies. Only the more significant links are shown, with weights represented by the line thickness and arrow size.  This model, with stochastic damage rates that can be implemented computationally \citep{Farrell2020}, can generate large synthetic populations that reflect observed high-dimensional health-states and mortality over the populations used to fit the model. See Figs.~3 and 4 below. Note that the full model only has $10$ nodes, representing just the observed health-states.}
  \label{fig:directed}
\end{figure}

The WNM can be used to generate synthetic individual health trajectories until mortality from any starting point. For example, in Fig.~3 we illustrate the joint distribution of the FI at death and the death age for four individuals with specific baseline deficits at their baseline ages (either 65 or 85 years). We see how both baseline age and deficits affect both the death age and the overall health at death. Intriguingly, the most likely FI at death for the older individuals is 1, which is well above typically reported $FI_{\mathrm{max}} \approx 0.7$ \citep{Stubbings2020}. Since we use only 10 deficits, the observed $FI_{\mathrm{max}}$ of this data is 1, but this does not explain the peak of the distribution at 1. The FI at death for individuals in this data is not observed, so this may be a prediction of a real effect related to terminal decline \citep{Stolz2020} -- however it may also partially be an artefact of over-simplified damage-rates within the model for individuals at higher FI (see discussion in \cite{Farrell2020}).

We can also look more closely at how health evolved up until death. These frailty trajectories are shown in Fig.~4, for individuals who died at their median predicted death age. This illustrates how baseline individual health and age can strongly affect subsequent health trajectories, and also how variable those trajectories are. 

While these synthetic populations resemble observed aging human populations, we are not yet able to robustly infer specific interactions between the observed health attributes. Instead, we find that many different networks are consistent with the observed data. We are currently developing a more generalized approach, using continuous-valued longitudinal datasets, to make individual predictions of aging trajectories and to infer a robust network of interactions on the level of blood biomarkers and functional disabilities.

\begin{figure}[t!] 
    \center{\includegraphics[width=0.45\textwidth]{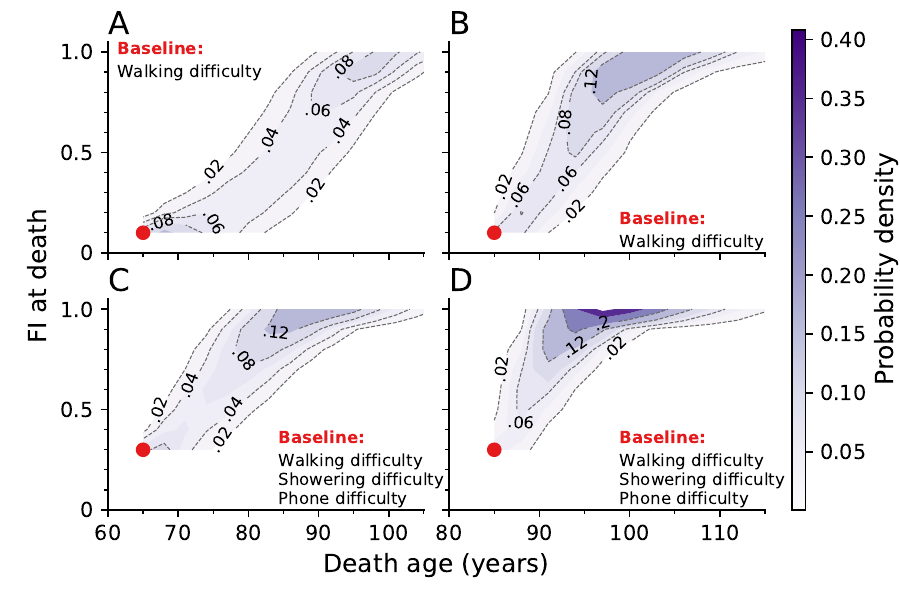}}
    \caption{Simulated joint distributions of FI at death and death age from the network model described in \citet{Farrell2020}, and illustrated in Fig.~2, given four individuals with the indicated initial baseline deficits (filled red points indicate their baseline age and FI). This demonstrates the capability of simulating populations of synthetic individuals, starting from different baseline conditions. The two columns show baseline ages of $65$ (A and C) and $85$ (B and D), the rows show different baseline FIs of $0.1$ (A and B) and $0.3$ (C and D), for an FI with 10 deficits.}
    \label{fig:Death}
\end{figure}

\section{Computational models}
All but the simplest models need to be implemented computationally. Computational models allow us to simulate and explore the quantitative consequences of various hypotheses. That is, since computational models require well-defined algorithms, they force us to make our assumptions explicitly. By varying those assumptions we can explore their consequences. This can clarify and illuminate possible mechanisms of aging.

Computational dynamical models can generate large synthetic populations of individuals with complete health trajectories and mortality. Such ``perfect'' data facilitates the systematic development of data analysis tools, including determining their statistical power for finite populations with missing data. More fundamentally, computational models also allow for a close examination of mechanisms at the population level: how does changing an assumption or a model parameter change the resulting health and mortality statistics of the population? The mechanisms behind the observed statistics of aging populations is a fundamental question of aging that can be answered, at least within the context of these model populations.

\begin{figure}[t!] 
    \center{\includegraphics[width=0.45 \textwidth]{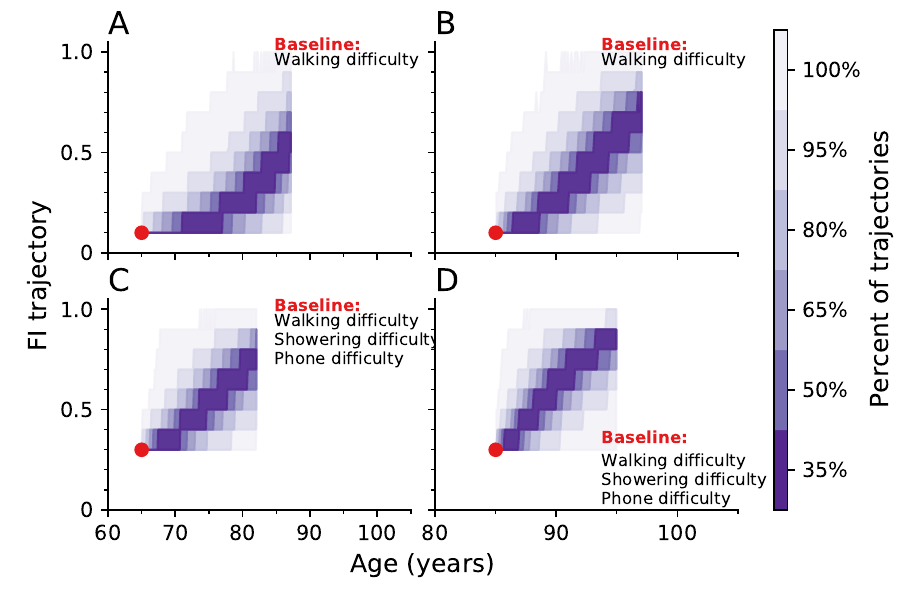}}
    \caption{Intervals of sampled stochastic FI trajectories vs age from the network model described in \citet{Farrell2020} and illustrated in Fig.~2, given given four individuals with the indicated initial baseline deficits. Trajectories are shown for samples with the median predicted death age, for the same individuals shown in Fig.~3. The purple shading indicates the percent of trajectories contained within the boundary of each region, similar to a confidence interval. The red circles indicates the baseline age and FI from which the simulation is started. The two columns show baseline ages of $65$ (A and C) and $85$ (B and D), the rows show different baseline FIs of $0.1$ (A and B) and $0.3$ (C and D), for a FI with 10 deficits.}
    \label{fig:Trajectories}
\end{figure}

The other side of the coin is that a significant \emph{disadvantage} of modelling is that every assumption needs to be explicitly built into the mathematical framework of the model, and every parameter needs to be determined. This is in contrast to observational studies of human or organismal populations, in which the assumptions and effective parameters are all implicitly included by biology. As a result, while natural observation naturally includes everything (including the ``kitchen sink''), a modelling approach typically only builds a minimal framework to address the mechanisms under consideration. Identifying an appropriate minimal framework, as well as specific mechanisms to consider, can benefit from qualitative approaches such as group model building \citep{Uleman2020}. Such approaches can help focus modelling efforts towards specific achievable goals, and will be particularly useful where effects across multiple scales are modelled (see \cite{Kenzie2017}) or more generally when faced with limited data.  

The constructive flavour of modelling lends itself to inter-organismal comparisons, since we can ask whether parameter tuning alone can explain differences between organisms or whether the structure of the aging model needs also to be changed.  Similar comparisons can be made between any distinct subpopulations of any one organism, including medically treated vs untreated populations or genetically distinct populations.

Structural changes in models should not be needed to accommodate small differences between populations. Conversely, structural differences between models will lead to distinctive effects that can be observed and therefore tested in observations of natural populations. However, testability will be challenging for mechanisms that are not already well characterized. For example, we expect that the effects of genetically heterogeneous human populations will eventually be important to characterize and include in models, but it will be hard to separate those effects from the intrinsic variability of the aging process. Nevertheless, a successful modelling framework should allow us to identify the statistical signatures of proposed mechanisms -- which  will facilitate subsequent testing. 

To paraphrase \citet{Box1976}, all models are at least partially wrong but some can nevertheless be useful. However, rather than just trying to be mostly right or fairly useful, models should  be improvable. This requires cycles of testing, development, and application to continually confront models with observable data. The benefit of this approach is that we can continually adjust our implicit or explicit assumptions to better and more usefully reflect emerging datasets.  

\subsection{Generalizability} 
Given the complexity of any organism, together with the complexity of the aging process, we can anticipate an enormous number of parameters required to tune complex models to fit a population. This tuning (also variously called fitting, learning, calibration, parameterization, or regression) is necessary if we want to do more than explore the qualitative consequences of a small set of model assumptions. Fitting the model to the data is necessary to make predictions for individuals, to compare populations, and to generate realistic synthetic populations.

We can distinguish between effective and fundamental parameters. Fundamental parameters are model independent, can be measured or derived with a variety of techniques, and are unchanged in different contexts. Effective parameters (like effective theories \citep{Transtrum2015}), on the other hand, cannot be precisely replicated in different contexts and cannot be derived from any fundamental assumptions. We expect that almost all parameters of models of aging will be effective parameters, i.e. they will be at least somewhat dependent on the choice of model.  Nevertheless, effective parameters should not be treated as arbitrary tuning-knobs of a model. More useful parameters will change less between studies, will be more interpretable, and will lead to better model predictions. Determining good model structure that facilitates useful parameterizations is an iterative process that is one goal of successful modelling. 

The technical details of fitting a model are well understood. Simple models can be hand-tuned to agree with population-level measures; however such simple models will not provide the best individual-level predictions. More sophisticated models can be fit with maximum likelihood, or other objective or ``loss'' functions, to obtain a model that best fits the available data. Bayesian approaches are also possible, where posterior distributions of parameters are obtained, rather than single point estimates.

An ideal data set would have a large homogeneous population, with complete, detailed health information that is longitudinally sampled frequently over individual lifetimes with uncensored mortality data. A computational model should be able to capture the important behavior exhibited in such data, so that it can then be used for individual predictions. Better computational models would provide better predictions. Are there other ways of distinguishing between such models? After all, even the best human data sets have small heterogeneous populations compared to national or global scales, with significant amounts of missing and censored data, and with irregular and infrequent longitudinal sampling with respect to the daily or weekly variability of our individual health status. 

The answer hinges on how generalizable the model is to different data sets. If the model fit to one dataset poorly generalizes to other datasets -- then the model has ``failed''. It has failed in a useful way \citep{Box1976} if we can expand the model in an interpretable way to accommodate both datasets. It has failed in a disappointing way if we cannot, and if we cannot understand why not.

Ideally, any successful model would be tested with as many different  datasets as possible. This builds confidence in the accuracy of the model, but conversely can uncover weaknesses in the model -- suggesting areas to improve. A significant limiting factor currently is the lack of multiple large, publically available, longitudinal datasets with similar health observations.

It will be exciting to ask whether we can also generalize models to different applications. For example, consider interventions in the health of individual organisms due to drugs, surgery, treatment, lifestyle, accidents, illness, or (in the case of model organisms) experimental manipulation. Can a model predict the outcomes of such interventions? Better predictions could be used to both improve individual treatment plans and manage the health-care of increasingly greying populations \citep{Harper2014}. 

Nevertheless, we might expect that any predictive model that is optimally tuned to predicting ``natural'' mortality or health outcomes will not be easily generalized to predict the outcomes of specific interventions. While part of this limitation naturally arises from the data used to train the model, some of this limitation will also come from the model structure itself, since computational models will not be able to accurately capture effects that are not allowed for in the model structure. Since the model structure itself may limit generalizability, some model structures (`types of models') will be better than others for this purpose. 

\begin{table*}[t!] 
\caption{Promising approaches for the modelling of aging.}
\begin{tabular}{|l|l|}
\hline
{\bf Longitudinal data}  &  \\[2pt]
\, \, Transitions between disease states at discrete times    &  \cite{Liu2015,Alaa2018}; \\ & \cite{Fisher2019, Walsh2020} \\[2pt] 
\, \, Irregularly timed observations & \cite{Rubanova2019, Schulam2015} \\ & \\[2pt]
\, \, Understanding covariate effects and interactions & \cite{Timonen2020} \\[5pt] \hline
{\bf Interpretable latent variables}    &  \cite{Avchaciov2020, Pierson2019}     \\[2pt] \hline
  {\bf Joint Models (survival)}   &  \cite{Lim2018}      \\[2pt] \hline
\end{tabular}
\label{Methods}
\end{table*}

\subsection{Underdetermined parameters, overfitting, and bias}
Any computational model of aging will also be at risk of underdetermined parameters, overfitting, and bias from the data sets used in training. These are generic problems of complex models. 

Underdetermined parameters are parameters of the model that are not well constrained by the available data, but are nevertheless important for model functioning. A loose analogy is that while \emph{a} hand may be needed for handwriting, \emph{which} hand is much less constrained. Given limited data, strong correlations often exist between parameters that each can range widely in magnitude -- this is the concept of \emph{sloppiness} \citep{Gutenkunst2007, Transtrum2015}. As a result, when considered individually these poorly constrained or ``slop\-py'' parameters have large uncertainties -- even when the model can still make robust predictions. 

To resolve this problem, we can focus on the predictions of complex models taking into account the uncertainties in underdetermined parameters, rather than the values of specific underdetermined parameters themselves \citep{Gutenkunst2007}. However, if direct interpretation of parameters is desired, determining specific parameter values can involve acquiring additional measurements, add\-ing assumptions to the model, or otherwise improving model \emph{identifiability} \citep{Chis2016}. Large clean datasets generated by computational models can be used to determine the types of observational data that would be needed to determine desired parameters. 

Overfitting is another generic problem with complex models with many parameters. Here, parameter values are fine-tuned to extract small improvements in fitting to the training data at the expense of good performance with new data. Overfitting is assessed by using dedicated training data and separate but comparable test data to assess model performance. Since fitting typically occurs through an iterative computational algorithm, overfitting can be minimized by simply stopping the fitting process when model performance on a held-out portion of the training data (the validation or development data) begins to decline.  

When test data is not comparable to training data then poor model performance can reflect poor generalizability due to limitations of the training data rather than due to overfitting. Some problem of generalizability arises in most training datasets, because they have biases in demographics (age distribution, sex, race), health-state of enrolled participants, medical treatment during the course of the study, or in any other possible category within the dataset. 

Modelling bias can also arise due to the structure of model not being able to account for all aspects of the data. For example, survivor bias \citep{Murphy2011} can be troublesome for models that do not capture mortality properly. When measured covariates have associations with mortality, the drop-out of individuals during the study due to mortality can bias the results. Models must account for survival effects, for example with joint longitudinal-survival models \citep{Hickey2016} that model health and survival together -- otherwise modelling efforts can erroneously try to accommodate survival effects within the disease progression itself.

\subsection{Specific Computational Approaches}
There are many possible approaches towards computational modelling. The most productive approaches will be determined by a combination of the background of the researcher, the problem at hand, and the data available. Agent-based modelling is one popular approach to managing multiple spatial and temporal scales in ecological systems \citep{Grimm2005} or in socially interacting populations \citep{Bonabeau2002}. However, given the large heterogeneous data-sets and  complexity inherent in aging organisms, we feel that Machine Learning (ML) techniques are particularly promising for aging research.  

\section{Machine Learning}
Machine Learning (ML) is a loosely-defined term for a collection of data-based models that are typically fit or ``trained'' with large high-dimensional data sets. Typical goals of ML approaches are classification (the most common application \citep{Domingos2012}, though not our focus here), regression, and generating synthetic samples with the same properties as the observed data. 

Neural networks  are often used in more sophisticated ML models \citep{Goodfellow2016}, as in deep learning \citep{LeCun2015}. Neural networks consist of layers of artificial neurons that each have many linearly combined input connections from previous layers, and many output connections to subsequent layers. All connection parameters for every neuron are trainable. Non-linear transformations in each neuron allow multiple layers (i.e. ``deep'' networks) to represent functions or relationships of arbitrary complexity \citep{Leshno1993, Raghu2017}.

Powerful neural networks have enormous numbers of parameters that must be trained for the network to represent a desired function. Neural networks are designed so that this training is computationally efficient. Overfitting can be a concern with so many parameters, and it is managed by careful use of regularization, which imposes restrictions on the parameters learned by the model. Test data, not used in training, is an important part of evaluating model performance and behavior.

Any unknown component of a model of aging can therefore be learned with a neural network, given sufficient training data. However, the researcher still needs to develop the overall structure of the model (i.e. how all the pieces glue together), choose appropriate neural network architectures, and manage the algorithms (and their ``hyper''-parameters) that train the model while avoiding overfitting. 

ML is rapidly developing, and new ML tools are easily learned and used after some expertise is gained. It is more challenging to achieve the goals of generalizability, where models perform well on data that is unlike the training set, and interpretability, where the mechanisms of the model can be understood and related to mechanisms in other model systems. Both of these goals are difficult because the flexibility of complex models required to achieve generalizability limits interpretability. Large data sets can help with generalizability by allowing the use of more complex models, but interpretability is an ongoing challenge \citep{Rudin2019}.

One strategy towards improving interpretability with ML is to develop models where knowledge of specific aging mechanisms are built into the model, while unknown components are learned with deep neural networks (using, e.g., differential equations to capture their behavior \citep{Rackauckas2020}). With this approach, encoding the knowledge we already have about the aging system effectively constrains the more general ML approach.  This maintains the flexibility that deep neural networks have to offer, allows the model to be trained with less data than a more general model, and adds interpretability to the model. 

Nevertheless, caution will be needed in accepting novel mechanisms just because they lead to better predictions or model performance. While we do expect that correct mechanisms will lead to better model behavior and generalizability, we also expect that powerful ML approaches may be able to perform reasonably well in spite of incorrect mechanisms. A critical approach will be called for, using data and predictions that can discriminate between putative mechanisms. 

\subsection{ML approaches in ageing research}
Several ML models have been developed specifically for aging. Pierson \emph{et al}. \citep{Pierson2019} developed a model that infers rates of aging for individuals that correlate with risk factors of aging, and that can be used to forecast future health. \citet{Avchaciov2020} developed a model that describes the aging of mice with an inferred dynamical frailty index, which correlates with both mortality and treatment effects. 

Similarly, machine learning  has already had  success in the estimation of biological age \citep{Hannum2013, Horvath2013, Levine2018, Lu2019}. With these models, many biological variables (e.g. DNA methylation levels) are reduced to a single estimate of biological age, which is found to be predictive of other health outcomes and mortality. The models used here are generally regularized linear models, due to the huge number of variables compared to the limited amount of data. As more data becomes available, more sophisticated techniques can be used for assessing biological age \citep{Putin2016, Pyrkov2018, Schultz2019, Zhavoronkov2019}. 

Nevertheless, models of biological age are not dynamical models -- they cannot simulate the future health trajectories of individuals, but only summarize and interpret the current health state. Furthermore, since reducing health to a single variable cannot capture multi-dimensional aspects of health, we believe that developing dynamical models that address longitudinal trajectories across multiple health dimensions is a promising direction for machine learning in aging research.

To forecast multi-dimensional health trajectories, existing machine learning approaches for modelling disease progression could be adapted to model aging progression. While many of these do not model mortality \citep{Schulam2015, Alaa2018, Fisher2019, Walsh2020}, joint longit\-udinal-survival models could be adapted for this purpose \citep{Lim2018}. A stochastic process model of aging has already been developed that models both health trajectories and mortality \citep{Yashin2007, Arbeev2011, Yashin2012, Arbeev2014}, but it has not yet been applied to high-dimensional datasets.

Given efficient algorithms for parameter determination (learning) together with flexible functional dependence (deep learning), we see great promise for ML approaches in the study of aging. Natural applications are filling in missing data, identifying natural subpopulations or categories of aging organisms, incorporating multiple heterogeneous data sources, and modelling the aging process itself as a stochastic dynamical process. 

In Table~\ref{Methods} we have listed some current work in the machine learning literature that we believe could be useful in applying machine learning to aging. While additional development of any existing techniques would be required for any specific problem, these approaches (and the references they cite) capture many useful ideas.

\section{The challenges and promise of aging models}
In the introduction, we listed four challenges of studying aging: how to better observe health in aging populations, how to better understand the mechanisms behind what we observe, how to better predict individual health, and how to better intervene in the aging process. We then highlighted particular challenges and promise of generalized computational models of aging in Table~1. Here, we provide more detail about the near-term challenges facing aging models -- together with some of the opportunities that make facing these challenges worthwhile.

\subsection{E-health, self-reported, and longitudinal data}
Focused population surveys are expensive. Large scale studies such as ELSA \citep{Steptoe2014} or NHANES \citep{NHANES2014} are limited to populations on the order of 10000 individuals. Even the impressive UK Biobank has less than one million individuals \citep{Sudlow2015}. In contrast, the use of electronic health records (EHR) \citep{Clegg2016} could eventually reach large fractions of national populations with lifetime longitudinal data. EHR is therefore an attractive source of data on the aging process. Similarly, individual health tracking through e.g. smart watches, or through self-reporting, could also reach large fractions of national populations. These developments will provide natural datasets with large populations that have lifetime longitudinal information. 

Significant biases are found in EHR data \citep{Vassy2018} and also in self-reported health data \citep{Zajacova2011, Gunasekara2012}. It will be difficult to explicitly account for these biases in order to reconcile EHR and self-reported data with corresponding national prospective studies from similar populations. Nevertheless, an immediate opportunity is to use this data for personalized health models and predictions. 

More generally, large-scale longitudinal data collection provides an opportunity for aging models to employ these data to better capture the aging process of individuals, including individual variability. Computational models of aging are well placed to make use of longitudinal data given the vast amount of data potentially available.  Natural questions include how much is gained by more frequent measurements, how to best handle variables observed at irregular time intervals and with varying degrees of missing observations, how to model mixtures of qualitative and quantitative measurements or of self-reported and molecular measures, and how to include individual health histories in individual health predictions.

\subsection{Defining and comparing populations}
Some \"{u}ber-model of aging might explicitly capture each aspect of individual variability, including a life-history of diet, lifestyle, injury, medication, and health-care. More realistically, most variability will first need to be captured implicitly within aging models through parameterization or model structure --  tuned for different natural subpopulations. Race \citep{Williams2005}, sex \citep{Gordon2018, Gordon2017}, socioeconomic position \citep{Knesebeck2007}, social vulnerability \citep{Wallace2015}, access to health-care \citep{Santana2000} or pensions \citep{Aguila2018}, rural/urban \citep{Yu2012}, and nationality, are all categories that have been studied by aging researchers. Chronic disease, genetic disorders, and certain patterns of multimorbidity or polypharmacy could also serve as natural categories. A challenge will be to reduce the significance of these natural categories for individuals by making the more of the implicit differences between the populations explicit -- allowing for better individualized study of aging and treatment. To be able to achieve this requires good data coverage across many subpopulations, but also good models that can characterize and model the differences.

Models of individual health with explicit context could then be used to generate synthetic populations that match measured or projected demographic information. This would be particularly useful for detailed projections of the effects of aging in population health.

\subsection{Multiple scales and subsystems}
Different physiological scales present exciting opportunities in aging research. For example, molecular data is appealing because it can be high throughput and low-bias. Nevertheless, outcomes at higher (functional) scales are typically of greater individual interest. One challenge is to identify interactions between scales, from molecular to behavioral, and to incorporate them in aging models \citep{Ferrucci2018, Mitnitski2019, Kuo2020}. Reliably bridging the scales, particular in light of patchwork individual data (over scale, over time, and over individual measures), is an important challenge. Understanding how different scales work is the essence of understanding the aging process. How does damage propagate from the molecular to  activities of daily living? Conversely, how do interventions of lifestyle or injuries propagate towards the molecular?

Diseases (such as Alzheimer's \citep{Fisher2019}), tissues (such as the brain \citep{Daunizeau2011}), or biological subfields (such as systems biology \citep{Dada2011}) each have distinct data-sets and modelling approaches. Particular aspects of the aging process, such as cellular senescence \citep{Karin2019}, can be similarly detailed.  We see two promising ways to combine specialized approaches with more generalized models of organismal aging. The first is to identify key summary measures from detailed models, and to train generalized models with accordingly pre-processed data. The second is to include generalized models as background aging processes within the more specialized models. Both should be explored, so that aging processes are more routinely combined with emerging biological, physiological, and medical models. 

\subsection{Predictions, treatment, and interventions}
Although individual predictions of health trajectories and mortality are natural goals for computational models of aging, a challenge is how to evaluate and judge the quality of the predictions, given the variety of different studies and possible outcomes. Evaluation of predictive quality is straight-forward retrospectively within the same dataset, by using separate training and test populations with either cross-sectional or longitudinal data with linked mortality. The determination of what quality of predictions are possible with what sort of data for individuals of a given age and health status will be important questions to answer.

Being able to predict the results of medical interventions (including medication), of illness or injury, or of life-style interventions such as exercise for aging individuals would be game-changing since it could be used to improve personalized treatment. Most individuals experience many such interventions over their lifetime, so these are implicitly and approximately included in models of national aging populations. Indeed, we assume that many such interventions are the origin of most national differences or differences within a national population over history.

A grand challenge will be to make many of these interventions explicit, particularly within models of individual health during aging. If successful, such explicit models will allow better individual prediction, better identification of intrinsic variability, and the ability to tailor or individualize interventions to better reflect individual priorities. To do this well we may need to include earlier data across individual life courses for large populations, including electronic health records and other longitudinal data. 

Current research is often focused on the ``diseases of aging'', such as Alzheimer's disease, cancer, and cardiovascular disease. While we are optimistic that complex computational models of aging can be applied to these conditions, doing so will require large-scale data for long-term outcomes of these specific diseases and possible interventions -- which may be difficult to obtain.

In the short term, there is a use for dynamical models in clinical trials since models can be used to generate large synthetic control populations \citep{Fisher2019, Walsh2020}. Models could synthetically create control arms that are better matched in age, sex, and original health status with respect to the treatment arm. Computational models may also be useful to explore and understand the effects of comorbidities and their treatments.

\section{Looking ahead}
Early modelling has been restricted to simple theoretical or statistical explorations of the aging process, through damage accumulation or regression models. Though this approach has limited ability to predict individual health, it has advanced our conceptual understanding of how aging could work.

More recently, various models have started to address observational data that includes the detailed health and mortality of large numbers of individuals, which we call ``networked'' models since they capture interactions between different aspects of individual health. Our work in this area has included explicit complex networks, but the networks can also be theoretical \citep{Yashin2012, Arbeev2016}, correlational \citep{Hidalgo2009, Roque2011, GarciaPena2019}, or implicit in the approach. Few models have addressed both individual health \emph{and} mortality, though these are now starting to emerge \citep{Farrell2020}.

Once multiple models with both health and mortality are developed, then the natural scientific selection of ``better'' models can proceed by confronting their simulated results with observed data. Natural measures of goodness of models include predictive quality, generalizability across different population demographics (including age and health, but also sex and chronic conditions), interpretability, and the ability to effectively and efficiently train with big heterogeneous data sets. The ability to efficiently and effectively predict future individual health trajectories will be revolutionary, particularly if models include the effects of injury and disease, or the benefit of various medical and pharmacological interventions. 

While observational data sets will only increase both in the number of individuals, in the number of physiological aspects of health that are reported, and in the frequency of longitudinal measurements, the amount of easy-available data available to train, test, and compare modelling approaches is still limited. Public ``challenge'' datasets could provide realistically imperfect but extensive longitudinal health data together with mortality statistics to allow for comparison between and improvement of modelling approaches. Providing raw data together with cleaned data is important, since improvements in data-cleaning \citep{VandenBroeck2005} can also lead to model improvement -- and computational pipelines of data-cleaning will be increasingly necessary for large population studies. 

We will never achieve a ``death-clock'' where we can precisely predict an individual's death, nor a health-calendar of precisely how their health will change as they age. Nevertheless, we may be able to classify and identify useful aging phenotypes, to obtain good predictions of individual health-trajectories and mortality, and to identify the most useful health interventions for a given individual. Because computational models can capture the effects of many interacting aspects of human physiology, they are also promising tools to use to help to address these questions. 

How computational models can and will be used will depend on how successful they become. We believe that they will lead to a deeper understanding of how aging works, both for human aging and for model organisms. By incorporating many different mechanistic effects within and between different organismal scales, computational modelling could reach towards an overarching, contingent, and quantitative theory of aging.  

More prosaically, computational models could help to control for the effects of different populations, or to improve national or regional comparisons of the determinants of health. We also expect that models will be able to capture the effects of various health interventions at the individual level. If models become sufficiently good, they would be able to help individuals to develop and adapt their personal health plans. We are hopeful.

\section*{Acknowledgements}
ADR thanks the Natural Sciences and Engineering Research Council (NSERC) for an operating Grant (RGPIN 2019-05888). KR has operational funding from the Canadian Institutes of Health Research (PJT-156114) and personal support from the Dalhousie Medical Research Foundation as the Kathryn Allen Weldon Professor of Alzheimer Research.

\bibliographystyle{spbasic}      
\bibliography{bib}
\end{document}